\documentstyle[12pt]{article}

\renewcommand{\ref}[1]{\raisebox{.6ex}{[#1]}}

\newcommand{\be}{\begin{equation}}
\newcommand{\ee}{\end{equation}}

\newcommand{\ba}{\begin{array}}
\newcommand{\ea}{\end{array}}

\begin{document}



\title{Friction on a Quantized Vortex in a Superfluid }

\author{X.-M. Zhu$^a$ and  P. Ao$^{b \dag}$  \\
       Department of Experimental$^a$/Theoretical$^b$ Physics \\
       Ume\aa{\ }University, S-901 87, Ume\aa, SWEDEN }

\maketitle


\begin{abstract}

We obtain the explicit expression of 
the friction on a moving quantized 
vortex, following the formulation of Thouless, Ao and Niu.
It is shown that the friction on a moving vortex
is sensitive to details but 
does not change the transverse force.
We provide a general thermodynamic interpretation for the mutual
independence of the transverse force and the friction.
The friction is evaluated for the
case of quasiparticle  contributions in a clean fermionic superfluid, 
showing a new feature of logarithmic divergence.

\noindent
PACS${\#}$s: 67.40.Vs; 47.37.+8 
 
\end{abstract}

 

In physical systems few qualities have generic values independent
of material details. When such cases occur, usually there are either
symmetry or topology reasons\cite{thouless}.
The controversy over dynamics of quantized vortices in
superfluids or superconductors\cite{brandt,review} 
may be partly  understood as due to the apparent conflict between
the exact results\cite{at,tan,az} with
the existence of various scattering processes  
in the system.  When a vortex moves,
in addition to the hydrodynamic Magnus force
which is proportional to the superfluid density,  
the coupling between the 
vortex to quasiparticles, phonons and impurities may
cause extra forces
in both transverse and longitudinal directions of the vortex motion.
Those possible non-vanishing forces
have always been the underlying assumption for phenomenological models
used in analyzing experiments\cite{brandt,donnelly}.

While the transverse force on a moving vortex is controversial,
the friction is poorly studied. The usual quoted friction
formula derived from a microscopic Hamiltonian is 
obtained through a
relaxation time approximation in force-force correlation functions. 
However, this type of procedure has been known to be wrong since 60's,
in the context of
obtaining the friction of electrons, i.e. resistivity.\cite{za}
The lack of explicit and correct
microscopic derivation of friction has two consequnces.
First, though the friction of vortex motion is
an important quantity in many experiments, it has not been received
its due attention. Second,
 the incomplete understanding of the friction has generated doubts 
on the exact result on the transverse force. 

A self-contained theory of both the friction and the transverse force
by a rigorous and elementary method is undoubtedly  desirable.
To provide such a calculation in an explicit and complete
manner is the main purpose of the present letter. 
The method we choose follows that of Ref. \cite{tan}.
In order to obtain the friction,
we use a limiting process
similar to the one in the transport theory 
 and in non-equilibiurm statistical mechanics. 
This limiting  process  has
only been mentioned
but not explicitly discussed in Ref.\cite{tan} due to 
its sensitivity to details.

In the following, after giving a general
expression for the friction, as an example,
we  evaluate the
friction for the case of a clean fermionic superfluid.
The resulting friction is completely new.
It comes 
from the off-diagonal potential scattering of the extended quasiparticles, 
and is stronger than the ohmic damping by a logarithmic diverging factor.
This new result illustrates the directness and usefulness of our
formulation, another purpose of the present paper.

We consider an isolated vortex in the superfluid whose position ${\bf r}_0$ 
is specified by a pinning potential.
The system is otherwise homogeneous and infinite.
There is no externally applied supercurrent or normal current.
The vortex is allowed to move slowly. Hence the system Hamiltonian $H$
contains a slow varying parameter ${\bf r}_0(t)$. The many body
wavefunction of the superfluid $| \Psi_\alpha(t) \rangle $
can be expanded in terms
of the instantaneous eigenvalues $E_\alpha({\bf r}_0 )$ and eigenstates
$|\psi_\alpha( {\bf r}_0 )\rangle $, for which we choose phases such that
$\langle\psi_\alpha|\dot\psi_\alpha \rangle=0$. Because of our
assumption of a  homogeneous superfluid, 
 $E_\alpha({\bf r}_0 )$ is independent of both ${\bf r}_0 $ and  time $t$.
With those considerations, the many-body wavefunction 
$| \Psi_\alpha(t) \rangle $ can be expressed by
\be
   |\Psi_\alpha(t)\rangle   =  e^{-iE_\alpha t/\hbar}
   |\psi_\alpha( {\bf r}_0 ) \rangle +\sum_{{\alpha'}\ne\alpha}
     a_{\alpha'}(t)
     e^{-i E_{\alpha'} t/\hbar} |\psi_{\alpha'}({\bf r}_0)\rangle \;, 
\ee
where, to the first order in velocity, $a_\alpha(t)=1$, and
\be
  a_{\alpha'}(t)= - \int^t_0 dt' \langle\psi_{\alpha'}|\dot\psi_\alpha \rangle
   e^{ i(E_{\alpha'} - E_\alpha ) t'/\hbar} \;.
\ee
%
%
Here $\nabla_0$ denotes the partial derivative with respect to
the position ${\bf r}_0$ of the pinning potential, the vortex position.
This gives the
expectation value of the force on the vortex as
\begin{eqnarray}
  {\bf F} & = & -\sum_\alpha f_\alpha \langle\Psi_\alpha|
   \nabla_0H|\Psi_\alpha \rangle  \nonumber  \\
      &  =  &
 -\sum_\alpha f_\alpha \langle\psi_\alpha|
   \nabla_0H|\psi_\alpha \rangle  \nonumber  \\
 & &  + \sum_{{\alpha'}\ne\alpha}f_\alpha 
    \langle\psi_\alpha({\bf r}_0(t) )|
    \nabla_0H|\psi_{\alpha'}({\bf r}_0(t) )\rangle \times \nonumber \\
 & &   \int^t_0 dt'  \langle\psi_{\alpha'}( {\bf r}_0(t') )|
   \dot{\psi}_\alpha({\bf r}_0(t') )\rangle 
  e^{ i(E_{\alpha'} - E_\alpha )(t'-t)/\hbar }     + c.c. \; , 
\end{eqnarray}
where $f_{\alpha}$ is the occupation probability of the state ${\alpha}$.
%
%
For a vortex moving with a small and uniform velocity ${\bf v}_V$,
\be
  |\dot{\psi}_\alpha( {\bf r}_0  )\rangle 
  ={\bf v}_V\cdot | \nabla_0 \psi_\alpha(  {\bf r}_0 )\rangle \;, 
\ee
The first term in the right hand side of Eq.(3)  is
independent of ${\bf v}_V$ and will be ignored.
In fact, it is zero because of the translational invariance regarding to the 
vortex position.
The integration in time
can be carried out directly. In addition, we also need to use  the following
relations, 
\be
   \langle\psi_\alpha|\nabla_0H|\psi_{\alpha'} \rangle
   = (E_{\alpha'}- E_\alpha )
    \langle \psi_\alpha|\nabla_0\psi_{\alpha'} \rangle 
   = (E_\alpha - E_{\alpha'} )
    \langle \nabla_0 \psi_\alpha| \psi_{\alpha'} \rangle  \; , 
\ee
which are obtained by taking gradient $ \nabla_0$ with respect to
$ H |\psi_{\alpha'} \rangle =  E_{\alpha'} |\psi_{\alpha'} \rangle$
and $   \langle \psi_\alpha|H =  E_\alpha \langle \psi_\alpha|$,
then multiplying  from left/right by 
$ \langle \psi_\alpha|$ or $ |\psi_{\alpha'} \rangle $ respectively.
After making simplifications by using Eq.(5), 
Eq.(3) gives
\[
  {\bf F}= \frac{1}{2}
  \sum_{ {\alpha'}\ne\alpha } \frac{ f_\alpha - f_{\alpha'} }
                                   { E_\alpha- E_{\alpha'} } 
  \frac{\sin(E_\alpha- E_{\alpha'} )t/\hbar }
       { ( E_\alpha- E_{\alpha'} )/\hbar }
  |\langle\psi_\alpha|\nabla_0H|\psi_{\alpha' } \rangle |^2 \; {\bf v}_V
\]
\be
  + i \hbar \sum_{ {\alpha'}\ne\alpha}
     f_\alpha \; \left(1-\cos(E_\alpha- E_{\alpha'})t/\hbar \right)
    \left\{ \left(
     \langle\psi_\alpha|\nabla_0\psi_{\alpha'} \rangle
    \times \langle \nabla_0\psi_{\alpha'}|\psi_\alpha\rangle \right)
     \cdot\hat{\bf z}\right\} \; 
      {\bf v}_V \times \hat{\bf z} \; , 
\ee
the longitudinal and transverse forces on a moving vortex.

To obtain the long time behavior 
of Eq.(6) requires a limiting procedure. There are two ways of
taking limiting sequences. A possible sequence is to take
the low frequency limit before the thermodynamic limit. 
As we will find out, the transverse force is independent of the
limiting process. Therefore such a limiting sequence is correct as long
as only the transverse force is concerned. In such a calculation, because 
the energy levels
have been treated as discrete ones, there is no friction.
In order to obtain friction,
we have to take the thermodynamic limit before the low frequency limit,
the correct limiting procedure.
For this purpose, we use the Laplace average\cite{kohn},
 $ \lim_{t\rightarrow\infty}{\bf F}(t)
  = \lim_{\epsilon\rightarrow 0^+}
   {\bf F}(\epsilon) $, where
$ {\bf F}(\epsilon)= \epsilon \int_{0}^{\infty} dt \; 
          {\bf F}(t) e^{-\epsilon t}$. 
 We will also use the identity $
  1/( \epsilon + i x)  =
  \pi\delta(x)-i P ( 1 /x ) $. 
For the transverse force ${\bf F}_{\perp}$, we have
\[ 
  \lim_{\epsilon\rightarrow 0^+}\epsilon
  \int_{0}^{\infty} dt  e^{ - \epsilon t} 
  \left(1-\cos(E_\alpha- E_{\alpha'})t/\hbar\right)
  = \lim_{\epsilon\rightarrow 0^+}
  \left(1 - \pi\delta( E_\alpha- E_{\alpha'} ) \epsilon \right) = 1\; ,
\]
because the summation over states in Eq.(6) is well behaved
regardless whether $ E_\alpha$ is discrete or continuous.
Hence
\be
  {\bf F}_{\perp} = i \hbar \sum_{{\alpha'}\ne\alpha}
     f_\alpha \; \left\{ \left(
     \langle\psi_\alpha|\nabla_0\psi_{\alpha' } \rangle
    \times \langle \nabla_0\psi_{\alpha'}|\psi_\alpha\rangle \right)
     \cdot\hat{\bf z}\right\} \; 
      {\bf v}_V \times \hat{\bf z} \; .
\ee 
This is precisely what has been obtained in Ref.\cite{tan}, where 
further calculations lead to
 $ {\bf F}_{\perp} = - h L\rho_{s} {\bf v}_V \times \hat{\bf z} $,
independent of details of the system. 
Here $\rho_s$ is the superfluid number density, $L$ the length of the vortex. 
The longitudinal force, friction, is given by 
${\bf F}_{\parallel}=-\eta  {\bf v}_V$ with
\be
 \eta = \frac{\pi }{2}
    \sum_{{\alpha'}\ne\alpha }   \hbar  
     \frac{f_\alpha - f_{\alpha'} }{ E_{\alpha'}- E_\alpha }
  \delta( E_\alpha- E_{\alpha'} )
   |\langle\psi_\alpha|\nabla_0H|\psi_{\alpha' } \rangle |^2  \, .
\ee
The coefficient of friction $\eta$ is determined by low energy 
excitations such as phonons, extended quasiparticles,
and localized quasiparticles when their discrete 
energy spectrum is smeared out by impurities. 
This expression 
is identical to the result in Ref.\cite{az} for the case 
of ohmic damping in the zero frequency limit. 
Eq.(8) will not pick up any superohmic contributions, 
and will give infinity for any subohmic contributions.
 We point out that while with the aid of Eq.(5)
the transverse force can be expressed only in terms of the wavefunction 
or the density matrix without explicitly refering to the Hamiltonian or 
its eigenvalues, as showing by Eq.(7), the longitudinal force, 
the friction, cannot, as showing by Eq.(8).
The explicit dependence on the Hamiltonian or its eigenvalues in Eq.(8) 
is the source of the sensitivity of friction to details of the system.

For the superfluid $^4$He, there is no full microscopic theory yet
\cite{donnelly}.
We will not attempt to evaluation of Eq.(8) for this superfluid here
because of the sensitivity of the friction to details.
The situation can be very different 
in the case of  the recent Bose-Einstein condensed systems, where
a well defined microscopic theory is supposed to be known.
Instead, as an example to illustrate the directness and usefulness, 
we evaluate Eq.(8) 
for the case of a homogeneous fermionic superfluid
using BCS theory with s-wave pairing. 
%
%
At finite temperatures the extended states above Fermi level,
the quasiparticles, are partially occupied. 
The vortex motion causes transitions between these states and gives rise to
friction.
The transitions between different single quasiparticle levels
$ \langle\psi_\alpha|\nabla_0H|\psi_{\alpha'} \rangle$
are considered here since they dominate the low energy process.
The quasiparticles
are described by the eigenstates, $u_\alpha $ and $v_\alpha $,
of the Bogoliubov-de Gennes equation. Their behaviors in the 
presence of a vortex has been well studied in Ref.\cite{bardeen}. 
We may take
\be
   |\psi_\alpha \rangle  = 
   \left( \begin{array}{c} u_\alpha (x) \\ v_\alpha (x) \end{array} 
       \right) = \frac{1}{\sqrt{L} }
    e^{ik_z z}e^{i\mu\theta + i\sigma_z\theta /2}\hat{f}(r) \; ,
\ee
with ${\bf r}$ measured from the vortex position, and $\theta$ the azimuthal
angle around the vortex.
In order to obtain an analytical form for the transition element,
we use a WKB solution for $\hat{f}(r)$,
\begin{eqnarray}
 \hat{f}(r)& = &  \frac{1}{ 2 \sqrt{ R\; r } }
   \left( \begin{array}{c} 
      [ 1 \pm \sqrt{ E^2 - |\Delta (r) |^2}/E ]^{1/2}  \\ { }
     [ 1 \mp \sqrt{E^2- |\Delta (r)|^2}/E ]^{1/2}   \end{array} 
       \right)\times \nonumber \\
 & &  \exp{\left\{i\int_{r_t}^r dr' \left( k_{\rho}^2 
      \frac{r'^2-r_t^2}{r'^2}\pm \frac{2m}{\hbar^2 } 
    \sqrt{ E^2- |\Delta (r')|^2}\right)^{1/2} \right\}} + c.c.  \; .
\end{eqnarray}
Here $k_{\rho}^2 = k_f^2 - k_z^2$,  $R$ is the radial size of the
system. This WKB solution 
is valid when $r$ is outside the classical turning point
$r_t = |\mu|/k_{\rho}$. 
Here $r_t $ is the impact parameter.
A WKB solution also exists inside
the turning point. However, because it approaches zero as $ 
(r k_{\rho})^{|\mu|} /|\mu|! $, the contribution to the transition elements
from this region is small, 
and will be set to zero.  
The transition elements are then given by
\begin{eqnarray}
   |\langle\psi_\alpha|\nabla_0H|\psi_{\alpha' } \rangle|^2
   & = & \left|\int dx ( u_{\alpha'}^{\ast}(x) \nabla_0 \Delta v_\alpha(x) + 
         v_{\alpha'}^{\ast}(x) \nabla_0 \Delta^{\ast} u_\alpha(x) ) \right|^2  
   \nonumber \\
   & = &  \left\{  \begin{array}{lc} 
    \frac{\Delta_\infty^4}{E^2R^2} \; 
    \delta_{k_{z1},k_{z2}} \delta_{\mu_1 ,\mu_2\pm1}
   \, , \; & |\mu| \leq \xi_0 k_{\rho} \\
   0 \, , & |\mu| >\xi_0 k_{\rho} \,  \end{array} \right. \; .
\end{eqnarray}
Here $\Delta_\infty$ is the value of $|\Delta (r)|$ 
far away from the vortex core.
%
%
Physically it means that if the classical quasiparticle trajectory 
is far away from
the vortex core, it will not contribute to the friction.
The summation over states in Eq.(8) is replaced by 
\[
  \sum_{{\alpha'}\ne\alpha}=
  \sum_{\mu_1 , \mu_2 , k_{z1}, k_{z2}} 
  \int dE_1 dE_2 
  \frac{E_1}{\sqrt{E_1^2-\Delta_\infty^2}}
  \frac{E_2}{\sqrt{E_2^2-\Delta_\infty^2}}
  \left(\frac{m}{\hbar^2 k_f}\right)^2\frac{R^2}{\pi^2} \, ,
\]
after considering the density of states. 

Substituting Eq.(11) into Eq.(8), using the quasiparticle distribution 
function $f_\alpha  = 1/(e^{\beta E_\alpha} + 1 )$, 
 the coefficient of friction is given by
\be
  \eta =
   \frac{Lm^2\xi_0 \Delta_\infty^4\beta}{8\pi^2 \hbar^3} 
   \int_{\Delta_0}^{\infty} dE 
   \frac{E^2}{E^2-\Delta_\infty^2}
   \frac{1}{E^2\cosh^2{(\beta E/2)}} \;  .
\ee
The integral in Eq.(12) diverges logarithmically. 
It implies that the spectral function
corresponding to the vortex-quasiparticle coupling is not
strictly ohmic but has an extra frequency factor proportional to
$\ln(\Delta_\infty/\hbar\omega )$\cite{cl,az}.
When  $\hbar\omega$ is not very small comparing to $\Delta_\infty$, 
which may be realized   when  close to T$_c$,
we can ignore the logarithmic divergence 
in Eq.(12) by using the density of states
for normal electrons to obtain a finite friction,
i.e. replacing  $E^2/(E^2-\Delta_\infty^2)$ with 1 in Eq.(12).
Close to  T$_c$, the friction approaches to zero the same way as 
$\Delta_\infty^2$, which is proportional to the superfluid density $\rho_s$.
When $ - \ln(\hbar\omega/\Delta_\infty)$ is large, we
need to use a more accurate
expression of vortex friction obtained in Ref. \cite{az}. 
Straightforward evaluation shows that in such a case
\be
   \eta = \frac{Lm^2\xi_0\Delta_\infty^3\beta}{16\pi^2\hbar^3 }
   \frac{1}{\cosh^2{(\beta \Delta_\infty /2)}}
   \ln (\Delta_\infty/ \hbar\omega_c ) \, .
\ee
Here $\omega_c$ is the low frequency cut-off.
It is determined by the size of the system for a single vortex, 
and by the inter-vortex distance for a vortex array.

It should be emphasized that the logarithmic
divergence comes from the interplaying between  the 
divergence in the density of states and the off-diagonal potential scattering.
We can consider a situation that we physically create a pinning center to
trap the  vortex and guide its motion. In such a case the vortex
has a diagonal potential. 
If the scattering is dominated by the diagonal potential, e.g., by the
pinning potential, additional factor coming from $|u_\alpha|^2 
- - - - |v_\alpha|^2$ will remove this logarithmic divergent.
This again shows the sensitivity of the friction to system details.

This extended state friction contribution exists 
for both clean and dirty superconductors
at finite temperatures. 
Close to the transition temperature, it scales linearly with the superfluid
density,
and is exponentially small when $ T << \Delta_\infty$.
For intermediate temperature $T \sim  \Delta_\infty$, 
using $\xi_0 \sim \hbar^2 k_F/m\Delta_\infty$ and
$N(0) = mk_F/\pi^2\hbar^2$,  $\eta 
\sim   L \hbar N(0) \Delta_\infty^2/k_BT $.  
For dirty superconductors additional contribution to the friction
arise from vortex core states. This contribution,
which remains finite at zero temperature,
was first discussed 
phenomenologically in Ref.\cite{bs}.
Eq.(12) also permits us to do a detailed 
microscopic calculation for the core contributions, 
as pointed out in Ref.\cite{az}.
Another application of the above results
may be the dynamics of the axisymmetric vortex 
in  $^3$He  B phase, where the 
Bogoliubov-de Gennes equation is essentially the same as the 
one in a s-wave superconductor\cite{vw}. 

As we have shown, the friction on moving vortex has
 certain interesting features which can be tested experimentally.
It is also clear that transverse force is not influenced by
the existence of a friction.  This is significant different
from  another familiar example,
the quantum  Hall system, where it is generally understood that the energy 
gap in the quasiparticle excitation spectrum 
stabilizes the topological structure, meanwhile  makes
the dissipation to vanish.
When the quantized Hall conductance occurs it 
associates with the vanishing of longitudinal voltage.
However, the total transverse force on a moving quantized vortex, which has a
topological significance, remains unchanged with finite friction.

There has been a great amount of effort in this subject, 
 because this is 
the only type of topological singularities which can be extensively studied 
experimentally, in both classical and quantum regimes. 
Some of the previous  work on 
vortex motion in fermionic superfluid has been discussed critically in
\cite{az,za}.
In bosonic superfluids, the situation of early work
 is more diverse\cite{donnelly}.
So far, the  experimental papers on superfluids and superconductors
claim support to 
different theoretical results\cite{zhu,mf,vinen,he3}.
We need to keep in mind 
that in general the real experimental situations
are far more complicated than ones usually 
considered in theoretical models. For example, 
in the analysis of Hall measurements in
superconductors, the vortex interaction and the pinnings  
are often ignored. If we use the
same assumption for longitudinal resistivity,
we would conclude that we should have flux flow 
for any current and at any temperature, which contradicts 
the longitudinal resistivity measurements\cite{brandt}.
Complications also exist in 
mutual friction measurements.
For example, the additional friction
caused by the relative
motion between the ends of the 
vortex lines and the diaphragm 
has not been considered in a recent 
experiment\cite{mf}. In addtion, because
these vortex lines are shorter than the viscous penetration depth of the
normal fluid, the boundary effect can also be important.
Conclusions on the transverse force drawed from such experimental data
should be taken with a due caution.

Before we conclude, we discuss  
the thermodynamic interpretation
 of our results.
We have shown that indeed a finite
friction does not change the transverse force.
To illustrate this, 
let us imagine a torus-shaped tank filled with superfluid.
The normal fluid is assumed to be at rest with the tank initially. 
After creating a vortex-antivortex pair, we 
apply an external force on the vortex to
let it wind around one of the circumference of the torus once with 
uniform velocity in time $t_{total}\rightarrow\infty$ before their
annihilation.
We consider the work done by the external force to the
system, which includes the superfluid, normal fluid and the tank(subtrate),
and the energy and entropy gain 
of the system.
The vortex is used as an intermediate step to find the forces.
The external force on  the vortex in the longitudinal direction
balances the force felt by the vortex due to its coupling to the normal
fluid. The  external force  on  the vortex in the transverse direction
balances the total force on the
superfluid, which is the summation of 
the Mangus force and a possible one from the normal fluid.

The superfluid velocity 
has a finite increase due to the phase change
caused by the vortex motion. Thus the kinetic energy
of the superfluid also
increases by a finite amount.
This kinetic energy increase is exactly equal to
the work done by the external force on the superfluid, 
if it is the Magnus force.
The frictions do not dissipate energy,
because the process is  quasi-static  in which the
normal fluid velocity is negligible.
Under such a condition, the
external force on the
superfluid in the transverse direction
needs to be at least
the magnitude of the Magnus force in order to provide the work
for the increase of kinetic energy of the superfluid.
The normal fluid and substrate 
cannot tranfer energy to the superfluid, which
carries no entropy,  by lowering their internal energies, 
because of the second law of thermodynamics\cite{callen}.

Our argument has not told us why the
total transverse force is not larger than
the Magnus force. If there is no entropy change in the
normal fluid, then the transverse force will not be larger than the
Magnus force because the energy increase in the system, carried by the
superfluid, not normal fluid, must 
equal to the work done by the external forces.
This is possible, because the superfluid carries zero entropy, and 
because our process is quasi-static, in which there is no
entropy generation within the system. There is also
no external heat bath to provide a heat flow.
The entropy stays constant
within the system.

In summary,  both the transverse force 
and friction
on a moving quantized
vortex has been obtained simultaneously within the framework
of Ref. \cite{tan}. The transverse force confirms  the
Berry phase results obtained with a trial many-body wavefunction\cite{at}.
New results on the friction are given for
a fermionic superfluid, which can be tested 
experimentally.

\noindent 
Acknowledgments {\ } {\ }
{ We thank David Thouless for useful comments. 
 This work was financially supported by Swedish NFR. }

{\ }

\noindent
$^{\dag}$Present address: Bartol Research Institute, University of 
Delaware, Newark, DE 19716, USA

\end{document}